\documentclass[a4paper,11pt]{article}
\usepackage{pos}

\title{Large $N$ conformal bootstrap and chirality}
\ShortTitle{Large $N$ conformal bootstrap and chirality}

\author*[a]{J.A. Gracey}

\affiliation[a]{ Theoretical Physics Division,
Department of Mathematical Sciences,
University of Liverpool, P.O. Box 147, Liverpool, L69 3BX,
United Kingdom \\}

\emailAdd{gracey@liverpool.ac.uk}

\abstract{We use the large $N$ critical point formalism to examine the 
interplay of supersymmetry and chiral symmetry on the location or otherwise of 
multiple zetas in the renormalization group functions of several related field
theories. In particular we determine that the anomalous dimension of the matter
field of the Lagrangian corresponding to the bosonic sector of the $O(N)$ 
Wess-Zumino is free of multiple zetas to $O(1/N^3)$.}

\FullConference{Loops and Legs in Quantum Field Theory (LL2026)\\
12-17, April, 2026\\
Bayreuth, Germany\\}

\renewcommand{\logo}{\relax}

\newcommand{\half}{\mbox{\small{$\frac{1}{2}$}}}
\newcommand{\threehalves}{\mbox{\small{$\frac{3}{2}$}}}
\newcommand{\partialslash}{\partial \! \! \! /}

\newcommand{\Nf}{N_{\!f}}

\begin{document}
\maketitle

\section{Introduction}

One topic of interest in quantum field theory over many years has been to 
understand the structure of perturbative renormalization group functions at
very high loop order. In particular cases it has been shown that multiple zeta 
values will appear in scalar field theories, \cite{1}. More recently Schnetz 
has demonstrated that new periods will occur at seven loops in scalar $\phi^4$ 
theory \cite{2}. Over the last four decades the general overview is that we now
understand the connection of multiple zeta values with loop order in a variety 
of core theories with different symmetries. Aside from flavour and colour group
properties, characterized by the generic parameter $N$, the main symmetries are
supersymmetry, gauge and chiral symmetry. The latter three are strictly only 
present in integer spacetime dimensions. When perturbative renormalization is 
carried out in dimensionally regularized extensions one has to address
potential inconsistencies that can arise in the regularized version before the 
limit to the critical dimension can be taken. To tackle such high perturbative 
orders requires a significant amount of resources. However a particular 
technique that can probe very high order perturbation theory is the $1/N$ 
expansion when $N$ is large and it effectively equates to carrying out 
perturbation theory in a related but different ordering of the underlying 
Feynman graphs. This method was used to study multiple zetas at high order in 
powers of $1/N$ but all orders in the coupling constant for scalar field 
theories in \cite{3} based on the results of \cite{4,5,6}.

In one application of the large $N$ method the $O(N)$ Wess-Zumino model,
\cite{7}, was examined in \cite{8} to third order. An interesting feature 
emerged which was that the multiple zetas that arose in the $O(N)$ scalar 
theories of \cite{3} were {\em absent} to the same order in $1/N$ in the 
Wess-Zumino model. That model is a supersymmetric theory of scalar chiral 
fields which is renormalizable in four dimensions. As it possesses two of the 
main continuous symmetries of that dimension it was not immediately clear which
of the two symmetries was driving the absence of multiple zetas to that order. 
Indeed it is already well-established that supersymmetry is related to the 
principle of maximal transcendentality, \cite{9}. In order to test which of the
two symmetries is at the heart of the multiple zeta absence one would need to 
study the theories that relate to the separate boson and fermion sectors of the
Wess-Zumino model. While the fermionic sector was examined at $O(1/N^3)$ in 
\cite{10} the bosonic sector has yet to be accessed to that order in $1/N$. We 
therefore report on progress with regard to this using the $1/N$ expansion. 
Here chirality means vertices are composed of directed edges with the flow 
either completely towards or away from the interaction point.

\section{Background}

For four dimensional theories where the matter fields reside in a symmetry 
group such as $O(N)$ or $SU(N)$ the $\beta$-function takes the general form
\begin{equation}
\beta(g) ~=~ \sum_{n=1}^\infty \sum_{r=0}^{n-1} b_{n-1 \, r} N^r g^n
\end{equation}
where $b_{00}$~$=$~$-$~$\epsilon$ in $d$~$=$~$4$~$-$~$2\epsilon$ dimensions and
$g$ is a generic coupling constant. For a core set of quantum field theories
such as Quantum Chromodynamics (QCD), $\phi^3$ and $\phi^4$ theory, the
Gross-Neveu-Yukawa and Wess-Zumino models they all share the property that
\begin{equation}
b_{11} ~\neq~ 0 ~~,~~ b_{22} ~=~ b_{33} ~=~ b_{44} ~=~ 
b_{55} ~=~ \ldots ~=~ 0 ~.
\end{equation}
We note that for QCD it is the number of quark flavours, $\Nf$, that these
restrictions apply to but not to the number of colours. The coefficients 
$b_{Li}$ are computed at successive $L$ loop Feynman graphs in perturbation 
theory. They can also be calculated in an expansion in a different small 
parameter which is $1/N$ if $N$ is large which corresponds to a reordering of 
terms
\begin{eqnarray}
\beta(g) &=& b_{00} g ~+~ b_{11} N g^2 \nonumber \\
&& +~ \left( b_{10} g^2 ~+~ b_{21} N g^3 ~+~ b_{32} N^2 g^4 ~+~
b_{43} N^3 g^5 ~+~ \ldots \right) \nonumber \\
&& +~ \left( b_{20} g^3 ~+~ b_{31} N g^4 ~+~ b_{42} N^2 g^5 ~+~
b_{53} N^3 g^6 ~+~ \ldots \right) ~+~ \ldots 
\label{betalargen}
\end{eqnarray}
When $g$~$=$~$O(\frac{1}{N})$ then successive lines in (\ref{betalargen}) are 
$O(1/N)$, $O(1/N^2)$ and $O(1/N^3)$ with respective coefficient sequences
$\{ b_{00}, b_{11} \}$, $\{ b_{p\,p-1} \}$ and $\{ b_{q\,q-2} \}$ with
$p$~$\geq$~$1$ and $q$~$\geq$~$2$. 

Alternatively defining
\begin{equation}
\beta(g) ~=~ -~ \epsilon g ~+~ ( b_{11} N + b_{10} ) g^2 ~+~
N g^2 \sum_{i=1}^\infty \frac{1}{N^i} \beta_i ( b_{11} N g )
\end{equation}
the functions, $\beta_i$ that contain the $b_{ij}$, can be accessed via the
critical exponent 
\begin{equation}
\omega(\epsilon) ~=~ -~ \frac{1}{2} \beta^\prime(g_c) ~\equiv~
\sum_{n=0}^\infty \frac{\omega_n(\epsilon)}{N^n}
\end{equation}
at the Wilson-Fisher fixed point formally giving for example
\begin{equation}
\beta(g) ~=~ -~ \epsilon g ~+~ ( b_{11} N + b_{10} ) g^2 ~-~
2 b_{11} N g^2 \int_0^{b_{11}Ng} \, d\xi \, \frac{\omega_1(\xi)}{\xi^2} ~+~
O \left( \frac{1}{N^2} \right) ~.
\end{equation}
Similar resummed expressions are straightforward to establish for other
renormalization group functions. The coefficients $b_{ij}$ are accessible via a
resummation of Feynman graphs where the canonical propagator exponent is 
replaced by functions of $d$ and $N$ and are determined from the properties of 
the Wilson-Fisher fixed point of the underlying universal critical theory.

The universal theory is defined by the kinetic terms of the matter field, which
lies in an $N$-tuplet of a flavour symmetry group, and the force-matter
interaction. For example the universal theory that contains $O(N)$ $\phi^4$ 
theory the Lagrangian is
\begin{equation}
L ~=~ \half \partial_\mu \phi^i \partial^\mu \phi^i ~+~
\half \sigma \phi^i \phi^i ~+~ f_\sigma(\sigma,\partial \sigma)
\end{equation}
where the critical coupling is regarded as unity in large $N$,
$1$~$\leq$~$i$~$\leq$~$N$ and $f_\sigma(\sigma,\partial \sigma)$ represents all
$\sigma$ dependent terms and self-interactions in even critical dimensions.
The scaling behaviour of the critical propagators are deduced by ensuring the
action remains dimensionless in $d$ dimensions giving
\begin{equation}
\langle \phi^i(x) \phi^j(0) \rangle ~=~
\frac{A \delta^{ij}}{(x^2)^\alpha} ~~,~~
\langle \sigma(x) \sigma(0) \rangle ~=~ \frac{B}{(x^2)^\beta}
\end{equation}
in coordinate space where $A$ and $B$ are $x$ independent amplitudes and the
full exponents are, with $d$~$=$~$2\mu$,
\begin{equation}
\alpha ~=~ \mu ~-~ 1 ~+~ \half \eta ~~~,~~~
\beta ~=~ 2 ~-~ \eta ~-~ \chi ~.
\label{albedef}
\end{equation}
Here $\eta$ is the anomalous dimension of the $\phi$ field and $\chi$ is the
anomalous dimension of the cubic vertex. The explicit expressions for the 
functions such as $\omega_1(\xi)$ are determined in $d$-dimensions by the large
$N$ critical point formalism of \cite{4,5,6}. Various critical exponents have 
been computed to all orders in $\epsilon$ to $O(1/N^2)$ or $O(1/N^3)$ in
\cite{5,6} and therefore perturbative coefficients are known at high loop order
before the corresponding loop computation has been carried out. More
importantly {\em all} orders insight into high order perturbative structures 
can be deduced.

For example the wave function exponent $\eta$ in $O(N)$ $\phi^4$ theory is 
known at $O(1/N^3)$ from the large $N$ conformal bootstrap method \cite{6}. It 
contains a term involving the function $I(\mu)$ defined by a two loop graph
\begin{equation}
I(\mu) ~=~ \left. \frac{\partial ~}{\partial \Delta} \ln \Pi(\Delta)
\right|_{\Delta ~\equiv~ 0}
\end{equation}
where $\Delta$ is the analytic regulator in coordinate space representation and
$\Pi(\Delta)$ is defined in Figure \ref{ipidef}. The $\epsilon$ expansion is 
known near four dimensions \cite{3,11} and contains multiple zeta values that 
agree with the seven loop $O(N)$ $\phi^4$ theory computations in \cite{2}. The 
same integral appears in $\eta$ at $O(1/N^3)$ in other theories such as 
Gross-Neveu-Yukawa, \cite{12,13}, and the supersymmetric nonlinear $\sigma$ 
model \cite{9}. By contrast for the $O(N)$ Wess-Zumino model $I(\mu)$ is 
absent, \cite{8}. In that instance the exponent $\eta$ was found at $O(1/N^3)$ 
without using the bootstrap formalism as it has not been constructed for 
superspace. The final expression only involves polygammas \cite{8}. More 
specifically the functions 
\begin{eqnarray}
\hat{\Psi}(\mu) &=& \psi(2\mu-3) + \psi(3-\mu) - \psi(\mu-1) - \psi(1)
\nonumber \\
\hat{\Theta}(\mu) &=& \psi^\prime(\mu-1) - \psi^\prime(1) \nonumber \\
\hat{\Phi}(\mu) &=& \psi^\prime(2\mu-3) - \psi^\prime(3-\mu) -
\psi^\prime(\mu-1) + \psi^\prime(1)
\nonumber \\
\hat{\Omega}(\mu) &=& \psi^{\prime\prime\prime}(\mu-1) -
\psi^{\prime\prime \prime}(1)
\end{eqnarray}
appear in the wave function exponent at $O(1/N^3)$.

\vspace{0.2cm}
{\begin{figure}[h]
\begin{center}
\includegraphics[width=9.00cm,height=2.25cm]{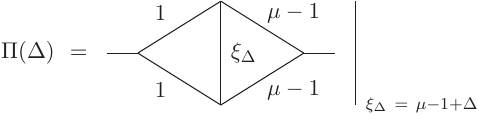} 
\end{center}
\caption{Two loop self-energy graph that defines $I(\mu)$.}
\label{ipidef}
\end{figure}}

The Wess-Zumino model is a cubic theory in chiral superfields $\Phi$ and
$\bar{\Phi}$ with the superspace interaction action, \cite{7},
\begin{eqnarray}
S_{\mbox{\footnotesize{int}}} &=& \frac{g}{3!} \int_x \,
\left[\int \!\! d^2 \theta ~ \Phi^3(x,\theta,\bar{\theta}) ~+~
\int \!\! d^2 \bar{\theta} ~ \bar{\Phi}^3(x,\theta,\bar{\theta}) \right] 
\label{wzact}
\end{eqnarray}
where the edges are all directed into or away from a vertex. The component 
Lagrangian contains scalar Yukawa interactions. A natural question arises as to
whether the absence of multiple zeta values is due to supersymmetry or chiral
symmetry or both. While the directed vertices means that only Feynman graphs 
with even edge subgraphs contribute to any Green's function, graphs that remain
could contain $I(\mu)$ which subsequently cancels in the sum. The way to 
resolve this is to study the corresponding scalar sector of (\ref{wzact}) using
the large $N$ conformal bootstrap.

\section{Large $N$ scalar chiral theory}

The scalar sector of the $O(N)$ Wess-Zumino model involves two doublets of
fields $\{\phi^i,\bar{\phi}^i\}$ and $\{\sigma,\bar{\sigma}\}$ where the 
universal Lagrangian for the large $N$ expansion is
\begin{equation}
L ~=~ \partial_\mu \bar{\phi}^i \partial^\mu \phi^i ~+~
\sigma \phi^i \phi^i ~+~ \bar{\sigma} \bar{\phi}^i \bar{\phi}^i ~+~
f_\sigma(\sigma,\bar{\sigma},\partial \sigma, \partial \bar{\sigma})
\label{lagscachi}
\end{equation}
with the final term representing the interactions that are active in even
critical dimensions. For the large $N$ conformal bootstrap formalism of
\cite{6} the coordinate space critical propagators are 
\begin{equation}
\langle \phi^i(x) \bar{\phi}^j(0) \rangle ~=~
\frac{A \delta^{ij}}{(x^2)^\alpha} ~~~,~~~
\langle \sigma(x) \bar{\sigma}(0) \rangle ~=~ \frac{B}{(x^2)^\beta}
\end{equation}
where $\alpha$ and $\beta$ are defined in (\ref{albedef}). We have applied the 
large $N$ bootstrap formalism of \cite{6} to deduce $\eta$ at $O(1/N^3)$ for 
(\ref{lagscachi}). In general the method involves analytic regularization with 
a parameter introduced by $\chi$~$\rightarrow$~$\chi$~$+$~$\Delta$; the 
spacetime dimension $d$~$=$~$2\mu$ is not the regulator. Briefly the method 
corresponds to perturbation theory in $\chi$. As the propagators include the 
anomalous dimensions no edges are decorated. Additionally all vertices are 
replaced by a conformal triangle, \cite{5,14,15}, meaning there are no vertex 
subgraphs. To the requisite order the relevant graphs and bootstrap vertex 
function is given in Figure \ref{scacfb} where the dot at a vertex indicates a 
conformal triangle and the double line denotes a $\sigma$ or $\bar{\sigma}$ 
field. So only two graphs contribute to the evaluation of $\eta$ at $O(1/N^3)$ 
which is denoted by $\eta_3$. 

{\begin{figure}[ht]
\begin{center}
\includegraphics[width=12.00cm,height=3.00cm]{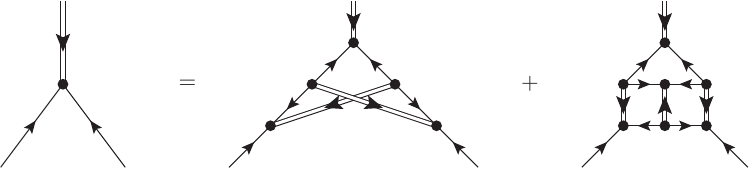}
\end{center}
\caption{Vertex consistency condition for the large $N$ bootstrap.}
\label{scacfb}
\end{figure}}

Applying the conformal bootstrap formalism of \cite{6} as well as that of 
\cite{4,5} we have computed all the core exponents for the fields and 
corrections to scaling in addition to $\eta_3$. In order to verify that the 
results are credible using {\sc Forcer} \cite{16,17,18} we have renormalized 
the relevant six dimensional Lagrangian to four loops which incorporates 
(\ref{lagscachi}) and is
\begin{equation}
L ~=~ \partial_\mu \bar{\phi}^i \partial^\mu \phi^i ~+~
\partial_\mu \bar{\sigma} \partial^\mu \sigma ~+~
\frac{g_1}{2} \bar{\sigma} \bar{\phi}^i \bar{\phi}^i ~+~
\frac{g_1}{2} \sigma \phi^i \phi^i ~+~
\frac{g_2}{6} \bar{\sigma}^3 ~+~ \frac{g_2}{6} \sigma^3 
\end{equation}
to determine the field anomalous dimensions and $\beta$-functions as well as 
the mixing matrix of all of the dimension two operators. Equipped with these
the respective critical exponents at the Wilson-Fisher fixed point are in full
agreement with the $\epsilon$ expansion of the $1/N$ exponents when
$\mu$~$=$~$3$~$-$~$\epsilon$, \cite{19}. Having checked that the perturbative 
and large $N$ expressions tally we find that in $d$-dimensions $\eta_3$ does 
not contain either $I(\mu)$ or $\hat{\Omega}(\mu)$. So like the Wess-Zumino 
model there are no multiple zetas to $O(1/N^3)$ in the $\phi^i$ field 
renormalization to all orders in perturbation theory. By way of illustration
\begin{equation}
\left. \frac{}{} \eta_3 \right|_{\mu=\frac{3}{2}} ~=~
\frac{128[494 - 15\pi^2]}{243\pi^6} ~~~,~~~
\left. \frac{}{} \eta_3 \right|_{\mu=\frac{5}{2}} ~=~ -~
\frac{65536[170503 + 375\pi^2]}{759375\pi^6}
\end{equation}
in odd dimensions showing that
$I(\threehalves)$~$=$~$2\ln 2$~$-$~$\frac{21}{\pi^2} \zeta_3$, \cite{6}, is 
absent.

To complete the investigation into whether multiple zetas are absent in 
$\eta_3$ in the Wess-Zumino model as a result of supersymmetry or due to 
chirality we recall the situation with the fermion sector. This was examined in
\cite{10} where the underlying universal large $N$ Lagrangian was used but 
based on the chiral Gross-Neveu model of \cite{20}. That Lagrangian is defined 
in terms of Majorana fermions meaning $\gamma^5$ was explicitly present in the
interaction. The application of the large $N$ conformal bootstrap computation 
then followed the parallel path as that of \cite{6}. However since the leading 
large $N$ term of $\chi$ vanished, in order to avoid potential singularities as
well as an intermediate $0/0$ issue, a non-abelian generalization of the 
universal Lagrangian was introduced, \cite{10}. This meant $\eta_3$ was 
computed as a function of Lie group Casimirs. It contained $I(\mu)$ but not 
$\hat{\Omega}(\mu)$. Moreover the $0/0$ issue could be traced to terms 
involving the ratio of purely non-abelian Casimirs which were non-problematic 
due to cancellations prior to taking the abelian limit. Thus in the final 
expression for $\eta_3$ in the chiral Gross-Neveu universality class the 
coefficient of $I(\mu)$ was zero. The non-abelian generalization could not be 
readily applied for (\ref{lagscachi}) since the consistency equation would have
involved all the topologies of equation (10) of \cite{6}. 

Having shown that chirality is a driving factor in the absence of multiple
zetas appearing at a particular order in large $N$ it is worth discussing other
aspects of this symmetry. First one interesting property is that the large $N$
critical exponents of the chiral Gross-Neveu model also precisely reproduce the
known perturbative four loop renormalization group functions of the four 
dimensional chiral XY model, \cite{21}. More specifically we repeated the 
computation of \cite{10} to three orders in $1/N$ but with the Weyl formulation
of the universal theory which is
\begin{eqnarray}
L^{\mbox{\footnotesize{WZf}}} &=& i \bar{\psi} \partialslash \psi ~+~
\sigma \psi \psi ~+~ \bar{\sigma} \bar{\psi} \bar{\psi} ~+~
f(\sigma,\bar{\sigma},\partial \sigma, \partial \bar{\sigma}) 
\label{lagcgnuniv}
\end{eqnarray}
and verified that the large $N$ critical exponents were in precise agreement 
with the $O(\epsilon^4)$ exponents derived from \cite{21}. This observation 
needs to be viewed in context. Perturbative computations are carried out in the
neighbourhood of the Gaussian fixed point. By contrast the large $N$ exponents 
are constructed at the Wilson-Fisher critical point where the underlying 
universal theory resides in all spacetime dimensions. It is only when the 
regularization is lifted in relation to the critical dimension do the Gaussian 
and Wilson-Fisher fixed points {\em collide}. In that limit the Lagrangian of 
the universal theory corresponds precisely to its strictly four dimensional 
counterpart. Similar features also hold for the large $\Nf$ properties of gauge
theories and the Wess-Zumino model.

\section{Discussion}

To summarize we have discussed large $N$ structural properties in different 
theories in respect of $\eta_3$. In particular the large $N$ analysis of the 
model corresponding to the bosonic sector of the Wess-Zumino model has been
completed at $O(1/N^3)$. This demonstrated that neither the scalar or fermion 
sector contain multiple zeta values at this order in the respective matter 
field wave function renormalization. To answer an earlier question the 
structural reason for this is the so-called chirality property which excludes 
graphs which have odd-edge subgraphs rather than supersymmetry. In related 
fermionic theories, such as the chiral Gross-Neveu model, this translates into 
the choice of Weyl fermions whence their algebraic properties circumvent the 
treatment of $\gamma^5$ in $d$-dimensions up to the loop and large $N$ orders 
studied here. This qualification follows in light of recent four loop 
computations in \cite{22} which suggest that care has to be taken when treating
$\gamma^5$ at six loops. If so the large $N$ critical point approach would not 
be exempt from having to find a parallel resolution which would appear first at 
$O(1/N^4)$. Another future direction to pursue is to see whether using the 
universal theory at the Wilson-Fisher fixed point with Weyl fermions for 
{\em perturbative} computations could provide an alternative or easier route to
treating $\gamma^5$ or supersymmetry within dimensional regularization.

\acknowledgments
This work was carried out with the support of the STFC Consolidated Grant
ST/X000699/1. For the purpose of open access, the author has applied a Creative
Commons Attribution (CC-BY) licence to any Author Accepted Manuscript version 
arising.

\end{document}